\documentclass[%
 reprint,
 amsmath,amssymb,
 aps,dvipdfmx
]{revtex4-1}
\usepackage{graphicx}% Include figure files
\usepackage{dcolumn}% Align table columns on decimal point
\usepackage{bm}% bold math
\usepackage{color}
\renewcommand{\a}{\alpha}
\renewcommand{\b}{\beta}
\newcommand{\g}{\gamma}
\newcommand{\bea}{\begin{eqnarray}}
\newcommand{\eea}{\end{eqnarray}}
\newcommand{\f}[2]{\frac{#1}{#2}}
\newcommand{\eq}{&=&}
\newcommand{\nn}{\nonumber \\ }

\renewcommand{\d}{\delta}
\newcommand{\area}{\int_{-\infty}^\infty }

\newcommand{\p}{\partial}
\newcommand{\pp}[2]{\f{\p #1}{\p #2}}
\newcommand{\s}{\sigma}
\newcommand{\sref}[1]{Eq. (\ref{#1})}

\newcommand{\om}{\omega}

\newcommand{\citeauthorname}[2]{{#1} {#2}}
\newcommand{\citebook}[4]{{#1} {\it #2} ({#3}, {#4}).}%{author}{bookname}{publisher}{year}
\newcommand{\citepaper}[4]{{#1} {#3} ({#4}).}%{author}{papername}{journal}{year}
\begin{document}

\preprint{APS/123-QED}

\title{Validation of the Replica Trick for Simple Models}% Force line breaks with \\

\author{Takashi Shinzato}
\email{takashi.shinzato@r.hit-u.ac.jp}
 \affiliation{
Mori Arinori Center for Higher Education and Global Mobility,
Hitotsubashi University, %Kunitachi, 
Tokyo, 1868601, Japan.}%Lines break automatically or can be forced with \\

\date{\today}% It is always \today, today,

\begin{abstract}
We discuss replica analytic continuation using several simple models 
in order to prove mathematically the validity of replica analysis, 
which is used in a wide range of fields related to large scale complex systems. 
While replica analysis consists of two analytical techniques, 
the replica trick (or replica analytic continuation) and the thermodynamical limit 
(and/or order parameter expansion), we focus our study on replica analytic continuation, 
which is the mathematical basis of the replica trick. We apply replica analysis to solve 
a variety of analytical models, and examine the properties of replica analytic continuation. 
Based on the positive results for these models we propose that replica analytic continuation 
is a robust procedure in replica analysis.
\begin{description}
\item[PACS number(s)]
{89.90.+n}, {75.50.Lk}
%\pacs{89.65.Gh}
\end{description}
\end{abstract}
\maketitle

\section{Introduction}
Replica analysis is a widely used analytic approach,
developed 
for the analysis of large scale complex systems 
in cross-disciplinary fields. Applications of replica analysis include the analysis of the physical properties of disordered magnetic 
materials, 
evaluation of the learning performance of neural networks, 
estimation of the transmission rate of 
channel coding and 
optimization of the investment risk of diversification investments 
\cite{Mezard1987,Nishimori,Mezard,EA,SK,KS,Amit1,Amit2,Gardner,%Shinzato2008,
Kabashima2004,Tanaka2002,Ciliberti,Shinzato-SA2015}. 
Replica analysis has a long history and has been discovered (and 
rediscovered) in several research fields \cite{Kac,Lin,Edwards1970,Edwards1971,Emery,Jasnow,Grinstein}.
In one of the pioneering works on replica analysis, 
Edwards et al. investigated the 
singular properties of disordered magnetic materials \cite{EA}, generally 
called spin glass. 
They first derived and used replica analysis to 
examine the physical properties of spin glass.
Based on 
the observation that 
2-body interactions between localized spins 
can be modeled by the Ruderma--Kittel--Kasuya--Yoshida interaction,
they assumed that 2-body interactions between localized spins 
can be randomly assigned and can be used to
{analyze} the singular physical properties of spin glass.
Subsequently, Sherrington et al. developed this spin glass model into
a fully connected spin glass model by applying the mean field 
approximation, 
since it is well known that solutions based on the mean field approximation are
the most rigorous \cite{SK,KS}.

Recently, following its success in the analysis of the physical properties of spin glass, replica analysis has been applied to the 
mathematical structures and
mathematical analogies of other 
large scale complex systems.
For instance, 
for the analysis of the learning performance of neural networks, 
Amit et al. applied replica analysis to 
the associative storage model and 
examined the mechanism for recalling stored memories, and 
evaluated the learning capacity of the associative storage model 
\cite{Amit1,Amit2}. 
Further, Gardner investigated 
the learning performance of perceptron, one of the most simplified neural network models,
using replica analysis, and found that 
it is consistent with the findings based on combinatorics 
in the previous work of Cover et al. \cite{Gardner,Cover,Venkatesh}.
As part of an evaluation of the transmission rate of communication technology, 
Kabashima et al. used a sparse connected spin glass model 
to resolve the decoding problem of the binary linear Low Density Parity Check code (LDPC) 
and guarantee that the communication performance of its linear code is 
close to the Shannon limit using replica analysis \cite{Kabashima2004}.
In addition, Tanaka
used the posterior probability as a type of perceptron learning
to resolve the maximizer of the posterior marginal estimation problem 
for
the third generation of radio communication technology
Code Division Multiple Access (CDMA) and
analyzed 
the bit error rate and channel capacity of CDMA 
using replica analysis \cite{Tanaka2002}.
To assess the risk of 
diversification investments, 
Ciliberti et al. 
dissected the minimal investment risks of the absolute deviation model and expected shortfall model 
in the zero temperature limit using replica analysis \cite{Ciliberti}. 
In a later work, 
Shinzato proved 
that 
the minimal investment risk and the investment concentration of the mean variance model 
are satisfied by the property of self-averaging 
using 
replica analysis and the Chernoff inequality \cite{Shinzato-SA2015}.

Replica analysis consists of two main steps, 
the replica trick operation
and 
the application of the thermodynamical limit. 
While several previous works have verified the effectiveness of replica 
analysis, 
in our own use of the replica trick, 
for convenience of analysis, 
we use a calculation technique where 
a number which is assumed to be an integer is replaced 
by a real or complex number in order to implement the limit operation, 
and for this reason replica analysis is not always guaranteed mathematically.
With this in mind, 
we compared the results estimated by replica analysis and other approaches 
such as the Markov chain Monte Carlo method and steepest descent 
method, 
and found that the effectiveness of replica analysis 
was partly validated {\cite{Shinzato-Kabashima}}.
However, 
in several previous works,
problems have arisen regarding negative entropy and 
the convexity of the order parameter expansion of free energy. 
To address these problems, the results of replica analysis were refined by the 1st step (or full step) replica symmetry breaking solution or Almeida--Thouless analysis,
an approach that can provide a partial fix, 
but does not necessarily verify the universal reliability of replica analysis \cite{Mezard1987,Nishimori,AT,Parisi1,Parisi2}.
With respect to 
these efforts to refine replica analysis, 
Ogure et. al.
analytically evaluated 
{the expectation of 
the complex power of 
the partition function} of the grand canonical discrete random energy model 
using complex analysis, and then 
compared 
three solutions obtained from replica analysis, 
the paramagnetic solution, 
ferromagnetic
solution, and 
first replica symmetry breaking solution (spin glass solution) \cite{Ogure1,Ogure2}.
Further, they 
derived the cases 
where the three solutions, composed of the $n$th moments of the partition function, are 
not analytic at $n=n_c\in(0,1)$ ($n$ is called the replica number),
and verified their results 
using numerical simulations.
In addition, Tanaka 
rearranged 
the problem implied in replica analysis 
using the framework of the moment problem, 
examining several counterexamples of the moment problem and 
summarizing the sufficient condition that 
the distribution of the partition function is uniquely 
determined from the moment sequences of the partition function \cite{Tanaka2007}.
Previous works have shown that 
we need to analyze the Helmholtz free energy in a large scale complex system 
in order to implement a quenched analysis of the system. 
Below 
we discuss the mathematical basis 
of 
the expectation of the logarithm of 
the partition function 
for the analysis of Helmholtz free energy. 
One of the disputed issues in replica analysis is 
the assumption that for an integer replica number there is
an analytical continuation to the real or complex number case,
as this is not always valid for replica numbers near $0$.
In order to validate the reliability of replica analysis mathematically, 
we need to examine the mathematical structure of replica analysis
 through the models that can be 
applied to the two distinct approaches for analyzing 
the $n$th moment of the partition function 
in a similar way to the argument discussed by Ogure et al.

In the present work, 
as the first step for resolving the validity of replica analysis
and 
rigorously interpreting the findings obtained from replica analysis,
we separate the two distinct analytical approaches, (1) 
the replica trick and (2) the thermodynamical limit (and/or the 
order parameter expansion),
and focus our study on the first of the techniques, the replica trick, or equivalently replica 
analytic continuation. Namely, our 
goal in this paper is to 
discuss 
the properties of replica analysis 
by examining replica analytic continuation.

This paper is organized as follows.
In the following section, 
we explain the analytical techniques of replica analysis, 
the replica trick and replica analytic continuation,
and consider the importance of the order of integration of two distinct variables. In Section \ref{sec3}
we test replica analytic continuation 
using several simple models 
and propose a conjecture regarding replica analytic continuation.
The final section is devoted to a
summary and plans for future work.

\section{From Replica Trick to Replica Analytic Continuation}

\subsection{Why Do We Need Replica Analysis?}
Based on several studies of 
the disordered magnetic material model, which is represented by 
the Edwards--Anderson model and the Sherrington--Kirkpatrik model \cite{EA,SK};
the neural network model, which is represented by 
the Hopfield model and perceptron model \cite{Amit1,Amit2,Gardner%,Shinzato2008
}; the channel coding model, which is represented by LDPC and CDMA \cite{Kabashima2004,Tanaka2002};
and the portfolio optimization problem, which is represented by the mean-variance 
model and absolute deviation model \cite{Ciliberti,Shinzato-SA2015},
it is well known that we can examine some {indicators which characterize} a system, for instance, the average magnetization, learning rate, and encoding rate, 
 by evaluating the Helmholtz free energy of the system $F=-\f{1}{\b}\log Z$ or its expectation 
$E[F]=-\f{1}{\b}E[\log Z]$, where $\b$ is the inverse temperature and 
 $Z$ is the partition function.
In several analyses of quenched disordered systems, 
replica analysis has been used constructively 
to assess the expectation of the Helmholtz free energy. 
However, the validity of several methods used in replica analysis have not been
guaranteed theoretically. For this, by comparing 
the results obtained by replica analysis and other approaches, such as the Markov chain Monte Carlo method and the steepest 
descent method, we can partly verify the validity of replica analysis. 
In this work, as a first step for confirming the effectiveness of replica analysis, we examine 
the replica trick, one of the two analytical approaches which 
constitute replica analysis.

\subsection{The Replica Trick}
In general, the replica trick consists of 
the following identity equation for any positive random variable $Z$,
\bea
\label{eq1}
\log Z\eq\lim_{n\to0}\f{Z^n-1}{n}.
\eea
If the random variable $Z$ is positive and its probability density function 
$P(Z)$ is known, using the replica trick and the expectation of the $n$th power of $Z$, $E[Z^n]=\int_0^\infty dZP(Z)Z^n$, 
the expectation of the logarithm of $Z$, $E[\log 
Z]=\int_0^\infty dZP(Z)\log Z$, is
estimated as follows:
\bea
\label{eq2}
E[\log Z]
\eq
\lim_{n\to0}
\left\{
\begin{array}{l}
\f{E[Z^n]-1}{n}\\
\f{1}{n}\log E[Z^n]\\
\pp{}{n}\log E[Z^n]
\end{array}
\right.,
\eea
where 
the three terms on the right hand side are consistent with each other,
which can be easily confirmed by using L'H$\hat{\rm o}$pital's rule. 
$n$ of $E[Z^n]$ is the 
replica number. However, 
a necessary condition for \sref{eq2} is that 
$n$ must be a real or complex number.
{In other words the result obtained for $n\in{\bf Z}$ is regarded as 
 $n\in{\bf R}$ and/or $n\in{\bf C}$. }
However, the validity of this assumption with respect to analytic continuation of the replica number is not always guaranteed.

It should be noted
that in general 
if we can accurately assess $E[Z^n]$ without approximation 
when the replica number is real or complex, then 
we can assess $E[\log Z]$ using \sref{eq2}, 
and therefore we are not limited to analyzing quenched disordered systems.
In this case, there is no question of the validity of 
the replica trick of \sref{eq2}.
We consider that 
the assumption that 
the result for $E[Z^n]$ in $n\in{\bf Z}$ can be solved 
using the thermodynamical limit and expanding the Helmholtz free energy on order 
parameters and its approximation can be 
regarded as the result for $E[Z^n]$ in $n\in{\bf R}$ or $n\in{\bf C}$ 
is the basis of the ambiguity in replica analysis.
\subsection{Two Types of Integral Assessments}
In the analysis of quenched disordered systems, which are
represented by the Sherrington--Kirkpatrick model in spin glass theory and 
the Hopfield model in 
associative memory, 
it is well-known that by analyzing the expectation of the Helmholtz free energy $F=-\f{1}{\b}\log Z$, 
$E[F]=-\f{1}{\b}E[\log Z]$, we can assess the typical behavior of the system.
To average the free energy, we use the replica trick of \sref{eq2} 
and analytically evaluate 
the $n$th power function of the partition function, 
$E[Z^n]$, in the right hand side of \sref{eq2}.
This requires the evaluation of two distinct integrals of 
the internal variable of the partition function, $S$, 
and the external variable of 
the partition function, 
$J$. 
Here, 
we consider the relationship between the order of the two integrations of $E[Z^n]$ and 
$n$.

We first consider 
the procedure first-S-next-J (hereafter abbreviated FSNJ) 
for which when evaluating $E[Z^n]$, 
the integral or sum of the thermodynamical variable or internal variable 
of the partition function $S$ is first implemented, 
followed by 
the integral or sum (or the expectation) over 
the configuration variable or external variable 
of the partition function 
$J$. Namely, 
for FSNJ
the expectation of $Z^n$, $E[Z^n]$, is derived as follows:
\bea
\label{eq3}
E[Z^n]\eq\sum_JP(J)\left(
\sum_Se^{-\b{\cal H}(S,J)}
\right)^n,
\eea
where ${\cal H}(S,J)$ is 
the Hamiltonian of the system, 
$Z=\sum_Se^{-\b{\cal H}(S,J)}$ is the partition function of the canonical ensemble of the inverse temperature 
$\b$, 
$\sum_J$ and $\sum_S$ are the integrals or sums over the 
whole configuration of 
$J$ and $S$, respectively, and 
$P(J)$ describes the probability function of 
the external variable $J$.
Since 
the integral or sum with respect to the internal variable $S$ is implemented first, 
we allow that 
the replica number $n$ in \sref{eq3}
is regarded as a real or complex number.

In the alternate procedure 
first-J-next-S (FJNS), 
the integral or sum of 
the external variable of 
the partition function is first implemented,
followed by 
the integral or sum of the internal variable.
That is, 
\bea
\label{eq4}
E[Z^n]=%\eq
\sum_{S_1}\sum_{S_2}\cdots
\sum_{S_n}
\left(
\sum_JP(J)
e^{-\b\sum_{a=1}^n{\cal H}(S_a,J)}
\right).\qquad
\eea
We first evaluate \sref{eq4} with respect to the external variable $J$ 
and next evaluate it with respect to the internal variables 
$S_1,\cdots,S_n$.
The intuitive 
advantage of FJNS is that 
for any $A,B>0$,
the factor expansion $(A+B)^{2.1}$ can be 
described by the infinite series expansion {and 
$(A+B)^2<(A+B)^{2.1}<(A+B)^3$ holds},
that is, since $(A+B)^n$
is described as a finite series expansion 
when the power number is an integer, 
we can evaluate each term of $(A+B)^2=A^2+2AB+B^2$ and 
$(A+B)^3=A^3+3A^2B+3AB^2+B^3$ and then approximately assess $(A+B)^{2.1}$ 
using a sort of false position method from the results of $(A+B)^2$ and 
$(A+B)^3$.
A peculiarity of this method is that if 
the accuracy of the approximation is reduced,
negative entropy and/or 
convexity of the order parameter expansion of the Helmholtz free energy may 
occur, casting doubt on the veracity of the replica analysis.

\begin{figure}[b]
\begin{center}
\begin{picture}(250,100)(30,0)
\put(60,-10){\vector(1,0){190}}
\put(100,-20){\vector(0,1){100}}
\put(100,-10){\circle*{5}}
\put(88,-26){$n=0$}
\put(128,-26){$n=2$}
\put(168,-26){$n=4$}
\put(108,2){$n=1$}
\put(148,2){$n=3$}
\put(188,2){$n=5$}
\put(120,-10){\circle*{5}}
\put(140,-10){\circle*{5}}
\put(160,-10){\circle*{5}}
\put(180,-10){\circle*{5}}
\put(200,-10){\circle*{5}}
\put(190,55){$n\in{\bf C}$}
\put(170,55){\circle{5}}

\put(120,75){$F\left(3.5+5\sqrt{-1}\right)=E\left[Z^{3.5+5\sqrt{-1}}\right]$}
\put(170,70){\vector(0,-1){10}}

\put(160,25){$G(4)=E[Z^{4}]$}
\put(180,20){\vector(0,-1){25}}

\put(210,-26){$n\in{\bf Z}$}
\end{picture}
\vspace{0.5cm}
\caption{
Relationship between $F(n)=E[Z^n],(n\in{\bf C})$ for FSNJ 
 and $G(n)=E[Z^n],(n\in{\bf Z})$ for FJNS on the complex plane.
\label{fig1}
}
\end{center}
\end{figure}

\subsection{Formulation of Replica Analytic Continuation}
We set the expectation of the $n$th power of the partition function derived 
by FSNJ as 
$F(n)=E[Z^n],(n\in{\bf C})$ and 
for FJNS as 
$G(n)=E[Z^n],(n\in{\bf Z})$, shown in Fig. \ref{fig1}. 
From \sref{eq4},
the assessment of $G(n)$ when $n\in{\bf Z}$ requires 
changing the order of the two integrations.

If we regard 
$G(n)$ as an analytical function with respect to $n$ and 
assume that 
$G(n)$ can also be described as an analytical function of 
$n$ when $n\not\in{\bf Z}$, then 
$F(n)=G(n)$ always holds in the case $n\in{\bf Z}$. 
We can then 
propose that this system demonstrates replica analytic 
continuation. By contrast, for 
$F(n)=G(n),(n\in{\bf Z})$ and 
$F(n)\ne G(n),(n\not\in{\bf Z})$, 
the system does not satisfy replica analytic continuation. For 
instance, 
when $n\in{\bf C}$, $F(n)=E[Z^n]=An^2+Bn+C+D\sin(2\pi n)$ {(where $D\ne0$)}
or 
$n\in{\bf Z}$, $G(n)=E[Z^n]=An^2+Bn+C$, since $F(n)\ne 
G(n),(n\not\in{\bf Z})$, replica analytic continuation does not apply.

If replica analytic continuation holds for the system under consideration, 
from the replica trick of 
\sref{eq2}, the expectation of the logarithm of the partition 
function, $E[\log Z]$, 
can be accurately analyzed.
We examine replica analytical continuation 
by applying it to physical or statistical models which can be 
evaluated for both the FSNJ and FJNS approaches, 
so that we can compare the results of the two approaches.
We have selected models for which the 
thermodynamical limit need not be applied,
so that we can test the accuracy of
replica analytical continuation alone, 
as part of our goal to focus on the replica trick.
Note that several previous works have used $G(n)=E[Z^n],(n\in{\bf 
Z})$, 
for which it is important to estimate $E[\log Z]$ accurately, 
and hence we need to consider the accuracy of $n$ in the neighborhood of $0$.
However, we do not try to evaluate the expectation of the logarithm of the 
partition function 
consistently, 
since we are interested mainly in 
the issue of replica analytic continuation, 
which is
not limited to the neighborhood $n=0$.

\section{Simple Examples\label{sec3}}

\subsection{Classical Harmonic Oscillator Model with Rayleigh Distribution
\label{sec3.1}}
First 
we discuss whether replica 
analytic continuation holds by applying the well known analytical model 
of the classical 
harmonic oscillator model. The Hamiltonian of the classical harmonic 
oscillator model, ${\cal H}(p,q)$, is defined as
\bea
{\cal H}(p,q)\eq\f{p^2}{2m}+\f{m\om^2}{2}q^2,
\eea
where $p$ and $q$ are the momentum and the position of the harmonic 
oscillator, respectively, $m$ is the mass of the particle and
$\om$ is the angular frequency of the harmonic oscillator. 
The partition 
function $Z$ is solved as follows:
\bea
Z\eq\f{1}{h}\area dpdqe^{-\b{\cal H}(p,q)}\nn
\eq\f{1}{\hbar\b\om},
\eea
using the Dirac constant $\hbar=\f{h}{2\pi}$.
Since the partition function of this model is analytically and easily 
evaluated,
it is suitable for discussing the mathematical structure of replica analytic continuation.

We assume that the angular frequency of the harmonic oscillator which 
models the vibration phenomena $\om$ has 
a Rayleigh distribution: 
\bea
P(\om)\eq
\left\{
\begin{array}{ll}
\om e^{-\f{\om^2}{2}}&\om>0\\
0&\text{otherwise}
\end{array}
\right..
\eea
From this, the FSNJ case with $n\in{\bf C}$ can be analyzed 
directly. The $n$th moment of $Z$, 
$F(n)=E[Z^n]$, is derived as
\bea
E[Z^n]
\eq
\int_0^\infty 
d\om P(\om)\left(\f{1}{\hbar\b\om}\right)^n
\nn
%d\om\om^{1-n}e^{-\f{\om^2}{2}}\nn
\eq\f{2^{-\f{n}{2}}}{(\hbar\b)^n}\Gamma
\left(1-\f{n}{2}\right).
\label{eq9}
\eea
Note that hereafter 
the Gamma function $\Gamma(s)=\int_0^\infty dt 
t^{s-1}e^{-t},(s\in{\bf C})$ is often used in our discussion \cite{Artin,Andrews,Levedev}.
It is a meromorphic 
function and has a pole at the origin and at negative integer numbers, 
so if $F(n)=E[Z^n]$ is described using the Gamma function, 
{
then we will omit $n\in{\bf C}$ which is related to a pole of Gamma 
in the discussion on replica analytic continuation.
}

For the FJNS case, since we can permutate the integral order, 
the $n$th moment of $Z$, $G(n)=E[Z^n]$, is calculated 
as
\bea
\label{eq10}
E[Z^n]
\eq\f{1}{h^n}
\int_0^\infty
d\om P(\om)
\area 
\prod_{a=1}^n
dp_adq_a\nn
&&
\exp
\left(
-\f{\b}{2m}
\sum_{a=1}^n
p_a^2
-\f{\b m\om^2}{2}
\sum_{a=1}^n
q_a^2
\right)\nn
\eq\f{1}{h^n}
\left(
\f{2\pi m}{\b}
\right)^{\f{n}{2}}
\area 
\prod_{a=1}^n
dq_a\f{1}{1+\b m
\sum_{a=1}^nq_a^2
}\nn
\eq
\left(
\f{2\pi}{h^2\b^2}
\right)^{\f{n}{2}}
\area 
\prod_{a=1}^n
dr_a\f{1}{1+
\sum_{a=1}^nr_a^2
}\nn
\eq\left(
\f{2\pi}{h^2\b^2}
\right)^{\f{n}{2}}
\f{2\pi^{\f{n}{2}}}{\Gamma
\left(\f{n}{2}\right)
}
\int_0^\infty dR\f{R^{n-1}}{1+R^2}.
\eea
In addition, we substitute the identity
\bea
2\int_0^\infty dR\f{R^{n-1}}{1+R^2}
\eq 2\int_0^{\f{\pi}{2}}d\theta\tan^{n-1}\theta\nn
\eq B\left(\f{n}{2},1-\f{n}{2}\right)\nn
\eq \f{\Gamma\left(\f{n}{2}\right)
\Gamma\left(1-\f{n}{2}\right)
}{\Gamma\left(\f{n}{2}+1-\f{n}{2}\right)},
\eea
into \sref{eq10}, giving 
\bea
E[Z^n]\eq\f{2^{-\f{n}{2}}}{\hbar^n\b^n}
\Gamma\left(1-\f{n}{2}\right),
\label{eq12}
\eea
where $r_a=q_a\sqrt{\b m}$, $R^2=\sum_{a=1}^nr_a^2$ and 
$B(p,q)=\f{\Gamma(p)\Gamma(q)}{\Gamma(p+q)}$ are used. 
Furthermore, with respect to the integral domain ${\cal 
D}=\left\{(r_1,\cdots,r_n)|\sum_{a=1}^nr_a^2<A^2\right\}$, from the 
identical integral
\bea
%&&
\int_{\cal D}
\prod_{a=1}^ndr_af\left(
\sum_{a=1}^nr_a^2
\right)%\nn
%\eq 
=
\f{2\pi^{\f{n}{2}}}{\Gamma\left(\f{n}{2}\right)}
\int_0^A dR R^{n-1}f(R^2),\qquad
\label{eq13}
\eea
the $n$-dimensional integral of $r_1,\cdots,r_n$ can be rewritten as an $n$-dimensional hyperspherical volume integral. From this, we find that 
$F(n)=E[Z^n],(n\in{\bf C})$ in \sref{eq9} derived by FSNJ is consistent 
with $G(n)=E[Z^n],(n\in{\bf Z})$ in \sref{eq12} by FJNS.
Namely, it is verified that replica analytic continuation holds for 
the classical harmonic oscillator.

\subsection{
\label{sec3.2}
Classical Harmonic Oscillator Model with 
 Positive Gaussian Distribution}
Similar to our discussion for the 
classical harmonic oscillator, 
we now discuss the case where the angular frequency 
$\om$ is drawn from a positive normal distribution, 
$P_\s(\om)$, defined as
\bea
\label{eq14}
P_\s(\om)\eq
\left\{
\begin{array}{ll}
\f{2}{\sqrt{2\pi\s^2}}e^{-\f{\om^2}{2\s^2}}&\om>0\\
0&\text{otherwise}
\end{array}
\right..
\eea
For FSNJ,
when $n\in{\bf C}$, the expectation of the $n$th power of 
$Z$, $F(n)=E[Z^n]$, can be assessed directly:
\bea
\label{eq15}
E[Z^n]\eq
\int_0^\infty d\om P_\s(\om)
\left(\f{1}{\hbar\b\om}\right)^n\nn
\eq
\f{2^{-\f{n}{2}}}{(\hbar\b\s)^n\sqrt{\pi}}
\Gamma\left(\f{1-n}{2}\right).
\eea
For FJNS, 
when $n\in{\bf Z}$, 
since we can permutate the integral order, 
$G(n)=E[Z^n]$ is assessed as
\bea
E[Z^n]
\eq\f{1}{h^n}\left(
\f{2\pi m}{\b}
\right)^{\f{n}{2}}
\area \prod_{a=1}^ndq_a\nn
&&
\int_0^\infty d\om\f{2}{\sqrt{2\pi\s^2}}
e^{-\f{\om^2}{2}
\left(
\f{1}{\s^2}+\b m\sum_{a=1}^nq_a^2
\right)
}\nn
\eq\f{(2\pi)^{\f{n}{2}}}{(h\b\s)^n}
\area \prod_{a=1}^ndr_a\f{1}{\sqrt{1+\sum_{a=1}^nr_a^2}}\nn
\eq\f{(2\pi)^{\f{n}{2}}}{(h\b\s)^n}\f{2\pi^{\f{n}{2}}}{\Gamma\left(\f{n}{2}\right)}
\int_0^\infty dR\f{R^{n-1}}{\sqrt{1+R^2}}\nn
\eq
\f{(2\pi)^{\f{n}{2}}}{(h\b\s)^n}\f{\pi^{\f{n}{2}}}{\Gamma\left(\f{n}{2}\right)}
B\left(\f{n}{2},\f{1-n}{2}\right)\nn
\eq\f{2^{-\f{n}{2}}}{(\hbar\b\s)^n\sqrt{\pi}}
\Gamma\left(\f{1-n}{2}\right).
\label{eq16}
\eea
Since $F(n)=E[Z^n],(n\in{\bf C})$ in \sref{eq15} is derived by FSNJ and 
$G(n)=E[Z^n],(n\in{\bf Z})$ in \sref{eq16} by FJNS are
also consistent, 
the replica analytic continuation is validated in this model where 
in \sref{eq16} 
$r_a=q_a\sqrt{\b m}$, $R^2=\sum_{a=1}^nr_a^2$, 
$B(p,q)=\f{\Gamma(p)\Gamma(q)}{\Gamma(p+q)}$ and \sref{eq13} are 
used.

\subsection{Classical Harmonic Oscillator Model with 
 Chi Distribution
\label{sec3.3}
}
We consider also
the case where $\om$ is 
distributed with a $\chi$ distribution, $P_{\chi}(\om)$, with degrees of freedom 
$s$, defined as
\bea
P_{\chi}(\om)\eq
\left\{
\begin{array}{ll}
\f{\om^{s-1}}{2^{\f{s-2}{2}}\Gamma\left(\f{s}{2}\right)}e^{-\f{\om^2}{2}}&\om>0\\
0&\text{otherwise}
\end{array}
\right..
\eea
For FSNJ, when $n\in{\bf C}$, $F(n)=E[Z^n]$, is evaluated as
\bea
\label{eq18}
E[Z^n]\eq
\int_0^\infty d\om P_{\chi}(\om)
\left(\f{1}{\hbar\b\om}\right)^n\nn
\eq
\f{2^{-\f{n}{2}}}{(\hbar\b)^n}
\f{\Gamma\left(\f{s-n}{2}\right)}
{\Gamma\left(\f{s}{2}\right)}.
\eea
For FJNS, 
when $n\in{\bf Z}$, $G(n)=E[Z^n]$ is calculated as
\bea
E[Z^n]
\eq
\f{1}{h^n}\left(
\f{2\pi m}{\b}
\right)^{\f{n}{2}}
\area \prod_{a=1}^ndq_a\nn
&&
\int_0^\infty d\om\f{\om^{s-1}}{2^{\f{s-2}{2}}
\Gamma\left(\f{s}{2}\right)
}
e^{-\f{\om^2}{2}
\left(
1+\b m
\sum_{a=1}^nq_a^2
\right)
}\nn
\eq\f{(2\pi)^{\f{n}{2}}}{(h\b)^n}
\area \prod_{a=1}^ndr_a\f{1}{\left(1+\sum_{a=1}^nr_a^2\right)^{\f{s}{2}}}\nn
\eq\f{(2\pi)^{\f{n}{2}}}{(h\b)^n}
\f{2\pi^{\f{n}{2}}}
{\Gamma\left(\f{n}{2}\right)}
\int_0^\infty dR\f{R^{n-1}}{(1+R^2)^{\f{s}{2}}}\nn
\eq\f{2^{-\f{n}{2}}}{(\hbar\b)^n{\Gamma\left(\f{n}{2}\right)}}B\left(
\f{n}{2},\f{s-n}{2}
\right)\nn
\eq
\f{2^{-\f{n}{2}}}{(\hbar\b)^n}
\f{\Gamma\left(\f{s-n}{2}\right)}
{\Gamma\left(\f{s}{2}\right)}.
\label{eq19}
\eea
$F(n)=E[Z^n],(n\in{\bf C})$ in \sref{eq18} analyzed by FSNJ is 
consistent with $G(n)=E[Z^n], (n\in{\bf Z})$ in \sref{eq19} analyzed by 
FJNS since these descriptions are consistent with 
$n\not\in{\bf Z}$, and thus 
replica analytic continuation may hold in this model. 
Note that this model includes the models in Subsections \ref{sec3.1} and \ref{sec3.2}
as special cases.
\subsection{Quantum Harmonic Oscillator Model with Exponential Distribution
\label{sec3.4}}
The models for which replica analytic continuation is satisfied are not limited 
to the classical harmonic oscillator model. Here, we discuss 
another commonly used analytical model, 
the quantum harmonic oscillator model. 
The Hamiltonian of the quantum harmonic oscillator model, 
${\cal H}(k)$, is defined as
\bea
{\cal H}(k)\eq\hbar\om\left(k+\f{1}{2}\right),
\eea
where $k$ is the quantum number and $\om$ is the positive angular frequency. 
In the case of a Boson, 
$k=0,1,2,3,\cdots$ 
and the partition function $Z$ in this model is calculated as
\bea
Z\eq\sum_{k=0}^\infty e^{-\b{\cal H}(k)}\nn
\eq\f{e^{-\f{\hbar\b\om}{2}}}
{1-e^{-\hbar\b\om}}.
\eea
Moreover, it is assumed that the angular frequency $\om$ is 
distributed with the exponential distribution $P_\mu(\om)$:
\bea
P_\mu(\om)\eq
\left\{
\begin{array}{ll}
\f{1}{\mu}e^{-\f{\om}{\mu}}&\om>0\\
0&\text{otherwise}
\end{array}
\right..
\eea
From this, for FSNJ,
when $n\in{\bf C}$,
$F(n)=E[Z^n]$
is evaluated as
\bea
E[Z^n]\eq\int_0^\infty d\om P_\mu(\om)
\f{e^{-\f{n\hbar\b\om}{2}}}
{(1-e^{-\hbar\b\om})^n}\nn
\eq\f{1}{\hbar\b\mu}
\int_0^1 dtt^{\f{n}{2}+\f{1}{\hbar\b\mu}-1}(1-t)^{1-n-1}\nn
\eq\f{1}{\hbar\b\mu}
B\left(\f{n}{2}+\f{1}{\hbar\b\mu},1-n\right),
\label{eq23}
\eea
where $t=e^{-\hbar\b\om}$ is used.

For FJNS, 
when $n\in{\bf Z}$, $G(n)=E[Z^n]$ is calculated as
\bea
\label{eq24}
E[Z^n]
\eq\sum_{\vec{k}=0}^\infty
\int_0^\infty 
d\om P_\mu(\om)
e^{-\hbar\b\om
\sum_{a=1}^n(k_a+\f{1}{2})
}\nn
\eq\sum_{\vec{k}=0}^\infty
\f{1}{\hbar\b\mu\left(\f{1}{\hbar\b\mu}+\f{n}{2}+\sum_{a=1}^nk_a\right)}\nn
\eq\sum_{m=0}^\infty\f{1}{\hbar\b\mu(\f{1}{\hbar\b\mu}+\f{n}{2}+m)}\left(
\begin{array}{c}
n-1+m\\
m
\end{array}
\right)\nn
\eq\f{1}{\hbar\b\mu\Gamma(n)}
\int_0^\infty ds
\int_0^\infty du u^{n-1}e^{-s(\f{n}{2}+\f{1}{\hbar\b\mu})-u+ue^{-s}}\nn
\eq
\f{1}{\hbar\b\mu}
\int_0^1 dtt^{\f{n}{2}+\f{1}{\hbar\b\mu}-1}(1-t)^{1-n-1}\nn
\eq\f{1}{\hbar\b\mu}
B\left(\f{n}{2}+\f{1}{\hbar\b\mu},1-n\right),
\eea
where $\vec{k}=(k_1,k_2,\cdots,k_n)^{\rm T}\in{\bf Z}^n$ and the Kronecker 
delta $\d(c,d)$ are used. Further 
\bea
\sum_{\vec{k}=0}^\infty
\d\left(m,\sum_{a=1}^nk_a\right)
\eq
\left(
\begin{array}{c}
n-1+m\\
m
\end{array}
\right),
\eea
and $t=e^{-s}$ are employed. From this, 
$F(n)=E[Z^n],(n\in{\bf C})$ in \sref{eq23} analyzed by 
FSNJ is consistent with 
$G(n)=E[Z^n],(n\in{\bf Z})$ in \sref{eq24} calculated by 
FJNS, indicating that 
replica analytic continuation 
also holds in this model.
\subsection{Random Field Ising Model\label{sec3.5}}
In this subsection, 
Fermi's quantum harmonic oscillator, the Fermion, with 
$k=0,1$, is discussed using the Ising model as the equivalent representation. For convenience, we here accept the Ising model, the 
Hamiltonian of which is defined as
\bea
{\cal H}(S)\eq -hS,
\eea
where $S(=\pm1)$ indicates the Ising spin and 
$h$ is the external magnetic field. Then, the partition function $Z$ 
is estimated easily as
\bea
Z\eq \sum_{S=\pm1}e^{-\b{\cal H}(S)}\nn
\eq 2\cosh \b h.
\eea
Furthermore, it is presumed that the external magnetic field $h$ is distributed with a 
Gaussian distribution with mean $\mu$ and variance $\s^2$, $P_{\mu,\s^2}(h)$:
\bea
P_{\mu,\s^2}(h)\eq\f{e^{-\f{(h-\mu)^2}{2\s^2}
}
}{\sqrt{2\pi\s^2}}.
\eea
Then, for FSNJ,
in the case of $n\in{\bf C}$,
$F(n)=E[Z^n]$ is solved as
\bea
E[Z^n]\eq\area dhP_{\mu,\s^2}(h)\left(2\cosh \b h\right)^n\nn
\eq \area Dx\left(2\cosh(\b\mu+\b\s x)\right)^n,
\label{eq29}
\eea
where $Dx=\f{dx}{\sqrt{2\pi}}e^{-\f{x^2}{2}}$
is employed.

For FJNS,
when $n\in{\bf Z}$, $G(n)=E[Z^n]$ is derived as 
\bea
E[Z^n]
\eq\sum_{\vec{S}\in\left\{\pm1\right\}^n}
\area dhP_{\mu,\s^2}(h)
e^{\b h\sum_{a=1}^nS_a}\nn
\eq\sum_{\vec{S}\in\left\{\pm1\right\}^n}
\exp
\left(
\b\mu\sum_{a=1}^nS_a
+\f{\b^2\s^2}{2}
\left(\sum_{a=1}^nS_a\right)^2
\right)\nn
\eq
\sum_{m=0}^ne^{\b\mu (2m-n)+\f{\b^2\s^2(2m-n)^2}{2}}
\left(
\begin{array}{c}
n\\
m
\end{array}
\right)\nn
\eq\area Dxe^{-n\b\mu-n\b\s x}
(1+e^{2\b\mu+2\b\s x})^n\nn
\eq \area Dx\left(2\cosh(\b\mu+\b\s x)\right)^n\label{eq30},
\eea
where $\vec{S}=(S_1,S_2,\cdots,S_n)^{\rm T}\in\left\{\pm1\right\}^n$ is 
used and the notation 
$\sum_{\vec{S}\in\left\{\pm1\right\}^n}$ means 
a sum over the whole configuration of $\vec{S}$. Furthermore, 
$\area Dx e^{\theta x}=e^{\f{\theta^2}{2}}$ and 
\bea
\sum_{\vec{S}\in\left\{\pm1\right\}^n}\d\left(2m-n,\sum_{a=1}^nS_a\right)
\eq\left(
\begin{array}{c}
n\\
m
\end{array}
\right),
\eea
are used. From \sref{eq29} and \sref{eq30},
$F(n)=E[Z^n],(n\in{\bf C})$ for FSNJ and $G(n)=E[Z^n],(n\in{\bf Z})$ 
for FJNS are equivalent, showing that 
replica analytic continuation holds for this model.

\subsection{Potential Energy for Near Earth Gravity\label{sec3.6}}
In this subsection, 
we discuss whether replica analytic continuation holds 
using a classical physics model that is not often used in 
statistical physics, the potential energy for near Earth gravity. For $g$, the gravitational acceleration,
the Hamiltonian ${\cal H}(x)$ is defined by the 
potential energy of the particle (its mass $m$) at position $x$, 
$mgx$, where the potential energy at origin is regarded as $0$: 
\bea
{\cal H}(x)\eq mgx.
\eea
We here consider the canonical ensemble of this model whose range is 
the set of non-negative real numbers.
The partition function of this model $Z$ is calculated as
\bea
Z\eq\f{1}{h}\int_0^\infty dxe^{-\b{\cal H}(x)}\nn
\eq\f{1}{h\b mg}.
\eea
Here it is assumed that the mass $m$ is randomly distributed with the Gamma distribution 
with shape parameter $\tau\in{\bf R}$, $P_\tau(m)$:
\bea
P_\tau(m)
\eq
\left\{
\begin{array}{ll}
\f{m^{\tau-1}e^{-m}}{\Gamma(\tau)}&m>0\\
0&\text{otherwise}
\end{array}
\right..
\eea
Using this distribution, for FSNJ, 
in the case of $n\in{\bf C}$, 
$F(n)=E[Z^n]$ is solved 
as
\bea
\label{eq35}
E[Z^n]\eq\int_0^\infty dm P_\tau(m)\left(\f{1}{h\b gm}\right)^n\nn
\eq\f{1}{(h\b g)^n}\f{\Gamma(\tau-n)}{\Gamma(\tau)}.
\eea

For FJNS, when $n\in{\bf Z}$,
$G(n)=E[Z^n]$ is described as
\bea
E[Z^n]\eq
\f{1}{h^n}
\int_0^\infty \prod_{a=1}^ndx_a
\int_0^\infty dmP_\tau(m)e^{-\b mg\sum_{a=1}^nx_a}\nn
\eq\f{1}{(h\b g)^n}
\int_0^\infty \prod_{a=1}^ndr_a
\f{1}{\left(1+\sum_{a=1}^nr_a\right)^\tau},
\eea
where $r_a=\b gx_a$ is used. Moreover, when $k>1$,
\bea
\int_0^\infty \f{dr}{(\a+r)^k}
\eq\f{1}{(k-1)\a^{k-1}},
\eea
is obtained. Using this formula, we can recursively solve $G(n)$ as
\bea
\label{eq38-1}
E[Z^n]\eq
\f{1}{(h\b g)^n}\f{1}{(\tau-1)}
\f{1}{(\tau-2)}\cdots
\f{1}{(\tau-l)}\nn
&&\int_0^\infty
\prod_{a=1}^{n-l}dr_a
\f{1}{\left(1+\sum_{a=1}^{n-l}r_a\right)^{\tau-l}}\nn
\eq
\f{1}{(h\b g)^n}\f{1}{(\tau-1)}
\f{1}{(\tau-2)}\cdots
\f{1}{(\tau-n)}\nn
\eq\f{1}{(h\b g)^n}\f{\Gamma(\tau-n)}{\Gamma(\tau)}.
\eea
$F(n)=E[Z^n],(n\in{\bf C})$ in \sref{eq35} analyzed by 
FSNJ and $G(n)=E[Z^n],(n\in{\bf Z})$ in \sref{eq38-1} analyzed by 
FJNS are consistent, and thus 
replica analytic continuation is guaranteed in this model.
\subsection{Moment of Chi-Squared Distribution\label{sec3.7}}
In this subsection 
we discuss whether replica analytic continuation holds 
for a model which is not used in physics, namely, where the random variable 
$Z$ is distributed with a $\chi^2$ distribution with degrees of freedom 
$m$, $P_{\chi^2(m)}(Z)$:
\bea
P_{\chi^2(m)}(Z)\eq
\left\{
\begin{array}{ll}
\f{1}{2^{\f{m}{2}}\Gamma\left(\f{m}{2}\right)}Z^{\f{m}{2}-1}e^{-\f{Z}{2}}&Z>0\\
0&\text{otherwise}
\end{array}
\right.,
\eea
where 
$Z$ of this $\chi^2$ distribution is the sum of 
the squares of $m$ random variables, $x_1,\cdots,x_m$, which are 
independently and identically drawn from the standard 
normalized distribution, that is, $Z=\sum_{i=1}^mx_i^2$. 
We regard the sum of squares $Z=\sum_{i=1}^mx_i^2$ to be the 
partition function.

For FSNJ, since the distribution of 
$Z$ is already known, 
that is, we can consider the thermodynamical average to be determined,
when $n\in{\bf C}$,
$F(n)=E[Z^n]$, is evaluated as
\bea
E[Z^n]\label{eq37}
\eq\int_0^\infty dZP_{\chi^2(m)}(Z)Z^n
\nn
\eq\f{2^n\Gamma\left(\f{m}{2}+n\right)}{\Gamma\left(\f{m}{2}\right)}.
\eea

For FJNS, since 
$Z$ is constructed from the sum of $m$ independently and 
identically distributed variables, 
that is, $Z=\sum_{i=1}^mx_i^2$, 
when $n\in{\bf Z}$, 
$G(n)=E[Z^n]$ is expanded as
\bea
E[Z^n]
\eq
\area \prod_{i=1}^mDx_i
\left(\sum_{i=1}^mx_i^2\right)^n\nn
\eq \area \prod_{i=1}^mDx_i
\sum_{\vec{k}=0}^n\f{n!}{\prod_{i=1}^m
k_i!
}\prod_{i=1}^mx_i^{2k_i}\d\left(n,\sum_{i=1}^mk_i\right)\nn
\eq 2^nn!\sum_{\vec{k}=0}^n
\d\left(n,\sum_{i=1}^mk_i\right)
\prod_{i=1}^m
\f{\Gamma
\left(k_i+\f{1}{2}\right)
}{\Gamma\left(\f{1}{2}\right)
\Gamma(k_i+1)
},\label{eq38}
\eea
where $\vec{k}=(k_1,\cdots,k_m)^{\rm 
T}\in\left\{0,1,2,\cdots,n\right\}^m$ and 
\bea
\area Dx x^{2k}\eq\f{2^k}{\Gamma\left(\f{1}{2}\right)}\Gamma\left(k+\f{1}{2}\right),
\eea
are used. 
We can verify that \sref{eq37} and \sref{eq38} are consistent by using mathematical induction.
For convenience, the right hand sides of \sref{eq37} and 
\sref{eq38} are rewritten as
\bea
F(m,n)\eq\f{2^n\Gamma\left(\f{m}{2}+n\right)}{\Gamma\left(\f{m}{2}\right)},\\
G(m,n)\eq2^nn!\sum_{\vec{k}=0}^n
\d\left(n,\sum_{i=1}^mk_i\right)
\prod_{i=1}^m
\f{\Gamma
\left(k_i+\f{1}{2}\right)
}{\Gamma\left(\f{1}{2}\right)
\Gamma(k_i+1)
}.\nn
\eea

When $m=1$, for any $n\in{\bf Z}$,
\bea
F(1,n)\eq\f{2^n\Gamma\left(n+\f{1}{2}\right)}
{\Gamma\left(\f{1}{2}\right)},\\
G(1,n)\eq
2^nn!\sum_{k_1=0}^n\d\left(n,k_1\right)\f{\Gamma\left(k_1+\f{1}{2}\right)}
{\Gamma\left(\f{1}{2}\right)
\Gamma(k_1+1)
}\nn
\eq 2^nn!\f{\Gamma\left(n+\f{1}{2}\right)}
{\Gamma\left(\f{1}{2}\right)
\Gamma(n+1)
}\nn
\eq 2^n\f{\Gamma\left(n+\f{1}{2}\right)}
{\Gamma\left(\f{1}{2}\right)
}.
\eea
This shows that $F(1,n)=G(1,n)$ holds.

Next, we set 
\bea
F(s,t)\eq\f{2^t\Gamma\left(\f{s}{2}+t\right)}{\Gamma\left(\f{s}{2}\right)},\\
G(s,t)\eq2^tt!\sum_{\vec{k}=0}^t
\d\left(t,\sum_{i=1}^sk_i\right)
\prod_{i=1}^s
\f{\Gamma
\left(k_i+\f{1}{2}\right)
}{\Gamma\left(\f{1}{2}\right)
\Gamma(k_i+1)
},\nn
\eea
and assume that 
$F(s,t)=G(s,t)$ holds when $m=s\in{\bf Z}$ for any $t\in{\bf Z}$. Then,
when $m=s+1$, 
\bea
&&G(s+1,t)\nn
\eq
2^tt!
\sum_{k_{s+1}=0}^t\sum_{k_s=0}^t
\sum_{k_{s-1}=0}^t\cdots
\sum_{k_1=0}^t\nn
&&
\d\left(t,\sum_{i=1}^sk_i+k_{s+1}\right)
\f{\Gamma\left(k_{s+1}+\f{1}{2}\right)}
{\Gamma\left(\f{1}{2}\right)
\Gamma\left(k_{s+1}+1\right)
}\nn
&&\prod_{i=1}^s\f{\Gamma\left(k_{i}+\f{1}{2}\right)}
{\Gamma\left(\f{1}{2}\right)
\Gamma\left(k_{i}+1\right)
}\nn
\eq 2^tt!
\sum_{k_{s+1}=0}^t
\f{\Gamma\left(k_{s+1}+\f{1}{2}\right)}
{\Gamma\left(\f{1}{2}\right)
\Gamma\left(k_{s+1}+1\right)
}\f{2^{k_{s+1}-t}}{\Gamma(t-k_{s+1}+1)}
\nn
&&
\left\{
2^{t-k_{s+1}}(t-k_{s+1})!
\sum_{k_s=0}^{t-k_{s+1}}
\sum_{k_{s-1}=0}^{t-k_{s+1}}\cdots
\sum_{k_1=0}^{t-k_{s+1}}\right.\nn
&&
\left.
\d\left(t-k_{s+1},\sum_{i=1}^sk_i\right)
\prod_{i=1}^s\f{\Gamma\left(k_{i}+\f{1}{2}\right)}
{\Gamma\left(\f{1}{2}\right)
\Gamma\left(k_{i}+1\right)
}
\right\}
\nn
\eq
2^tt!
\sum_{k_{s+1}=0}^t
\f{\Gamma\left(k_{s+1}+\f{1}{2}\right)}
{\Gamma\left(\f{1}{2}\right)
\Gamma\left(k_{s+1}+1\right)
}\f{2^{k_{s+1}-t}}{\Gamma(t-k_{s+1}+1)}
\nn
&&G(s,t-k_{s+1}).
\eea
Further, from the assumption of mathematical induction,
since $G(s,t-k_{s+1})=F(s,t-k_{s+1})$ holds, we have 
\bea
&&G(s+1,t)\nn
\eq
2^tt!
\sum_{k_{s+1}=0}^t
\f{\Gamma\left(k_{s+1}+\f{1}{2}\right)}
{\Gamma\left(\f{1}{2}\right)
\Gamma\left(k_{s+1}+1\right)
}\f{2^{k_{s+1}-t}}{\Gamma(t-k_{s+1}+1)}
\nn
&&
2^{t-k_{s+1}}\f{\Gamma\left(\f{s}{2}+t-k_{s+1}\right)}
{\Gamma\left(\f{s}{2}\right)}\nn
\eq\f{2^t}{\Gamma\left(\f{1}{2}\right)
\Gamma\left(\f{s}{2}\right)
}\sum_{k_{s+1}=0}^t
\f{t!}{k_{s+1}!(t-k_{s+1})!}\nn
&&
\Gamma\left(k_{s+1}+\f{1}{2}\right)
\Gamma\left(t-k_{s+1}+\f{s}{2}\right)\nn
\eq
\f{2^t}{\Gamma\left(\f{1}{2}\right)
\Gamma\left(\f{s}{2}\right)
}\sum_{k=0}^t
\f{t!}{k!(t-k)!}\nn
&&
\int_0^\infty du u^{k+\f{1}{2}-1}e^{-u}
\int_0^\infty dv v^{t-k+\f{s}{2}-1}e^{-v}\nn
\eq
\f{2^t}{\Gamma\left(\f{1}{2}\right)
\Gamma\left(\f{s}{2}\right)
}
\int_0^\infty duu^{\f{1}{2}-1}e^{-u}
\int_0^\infty dvv^{\f{s}{2}-1}e^{-v}\nn
&&\sum_{k=0}^t
\f{t!}{k!(t-k)!}u^kv^{t-k}\nn
\eq 2^t
\int_0^\infty du\f{u^{\f{1}{2}-1}e^{-u}}{\Gamma\left(\f{1}{2}\right)}
\int_0^\infty dv\f{v^{\f{s}{2}-1}e^{-v}}{\Gamma\left(\f{s}{2}\right)}\nn
&&(u+v)^t.
\eea
Setting $u=\f{x}{2}$ and $v=\f{y}{2}$ we obtain
\bea
&&G(s+1,t)\nn
\eq 2^t
\int_0^\infty dx\f{x^{\f{1}{2}-1}e^{-\f{x}{2}}}{\Gamma\left(\f{1}{2}\right)2^{\f{1}{2}}}
\int_0^\infty dy\f{y^{\f{s}{2}-1}e^{-\f{y}{2}}}{\Gamma\left(\f{s}{2}\right)2^{\f{s}{2}}}
\left(\f{x+y}{2}\right)^t\nn
\eq\int_0^\infty dxP_{\chi^2(1)}(x)
\int_0^\infty dyP_{\chi^2(s)}(y)(x+y)^t\nn
\eq
\int_0^\infty dzP_{\chi^2(s+1)}(z)z^t\nn
\eq\f{2^t\Gamma\left(\f{s+1}{2}+t\right)}
{\Gamma\left(\f{s+1}{2}\right)}\nn
\eq F(s+1,t).
\eea
Thus, $F(s+1,t)=G(s+1,t)$ holds. Note that we use 
the property that the sum of the random variable $x$, which is drawn from the $\chi^2$ distribution with $1$ degree of freedom, 
and 
the random variable $y$, which is drawn from the $\chi^2$ distribution with 
degrees of freedom $s$, so that $z=x+y$ is 
distributed with a $\chi^2$ distribution with degrees of freedom $s+1$.
From this induction, it is verified that $F(m,n)=G(m,n)$ holds. Finally, 
since both descriptions, $F(n)=E[Z^n],(n\in{\bf C})$ for 
FSNJ and $G(n)=E[Z^n],(n\in{\bf Z})$ for FJNS, are also consistent 
with each other,
we have shown that replica analytic continuation holds
 in this model.

\subsection{Other Toy Models\label{sec3.8}}
Lastly, we introduce other toy models which satisfy replica 
analytic continuation:
\begin{enumerate}
\item From the partition function $Z=\area Dxe^{x\om}=e^{\f{\om^2}{2}}$ and 
$P_\s(\om)$ in \sref{eq14}, $F(n)=E[Z^n]=(1-n\s^2)^{-\f{1}{2}}$ is 
      obtained. 
Since 
$x_a\sim N(0,1),(a=1,\cdots,n)$, it is well known that 
$s=\s\sum_{a=1}^nx_a$ is distributed with a Gaussian distribution with 
      mean $0$ and variance $n\s^2$. Using $s$,
we can also easily analyze $G(n)$ for FJNS.
\item We consider a random variable $Z$ from the $F$ distribution 
      with degrees of freedom $s$ and $t$:
\bea
P(Z)\eq
\left\{
\begin{array}{ll}
\f{s^{\f{s}{2}}t^{\f{t}{2}}
}{B\left(\f{s}{2},\f{t}{2}\right)}
\f{Z^{\f{s-2}{2}}}{(sZ+t)^{\f{s+t}{2}}}
&Z>0\\
0&Z\le0
\end{array}
\right..
\eea
Then, $F(n)=E[Z^n],(n\in{\bf C})$ is calculated as follows. 
\bea
E[Z^n]\eq\left(\f{t}{s}\right)^n\f{\Gamma\left(\f{s}{2}+n\right)
\Gamma\left(\f{t}{2}-n\right)
}{\Gamma\left(\f{s}{2}\right)
\Gamma\left(\f{t}{2}\right)}
\eea
We can regard the random variable $Z$ as being the ratio between 
      the squared average of 
$s$ random variables $x_k,(k=1,\cdots,s)$ which are independently and 
      identically distributed with a standard normal distribution, and 
the squared average of $t$ random variables $y_l,(l=1,\cdots,t)$ which are also 
      independently and identically distributed with 
a standard normal distribution as follows:
\bea
Z\eq\f{\f{1}{s}\sum_{k=1}^sx_k^2}
{\f{1}{t}\sum_{l=1}^ty_l^2}.
\eea
Thus, $G(n)$ is also easily solved using the mathematical induction in Subsection \ref{sec3.7}.
\end{enumerate}

\subsection{Discussion and Open Problems}
As stated above, in this study 
we have examined 
replica analytic continuation, the mathematical basis behind the replica trick, one of the two steps that comprise replica analysis. Some of the concerns about the reliability of replica analysis are based on doubts about the accuracy of the replica trick.
When applying replica analysis,
both the replica trick and the thermodynamical limit 
are used, and hence it can be hard to unravel any inaccuracies in replica analysis.
Therefore, we have applied replica analysis to 
some physical or statistical models for which the 
thermodynamical limit need not be applied,
so that we can test the accuracy of
replica analytical continuation alone, 
thereby definitively addressing part of the basis of replica analysis.

In this paper, although we considered only a few examples due to the limitations of space, 
it is clear from the results of the models for which 
we can apply 
both analytical procedures, FSNJ and FJNS,
that we can assess $F(n)=E[Z^n],(n\in{\bf C})$
and $G(n)=E[Z^n],(n\in{\bf Z})$ without approximation 
and can verify that 
replica analytic continuation is guaranteed in these models. 
We note that a more complete analysis would examine a counterexample to the case of replica analytical continuation 
in mathematical models which can be analyzed without approximation.
Thus, 
while the analysis in this paper does not provide a universal mathematical proof of replica analysis, the positive examples detailed in this 
paper lead us to the following conjecture:
\begin{description}
\item[Conjecture on Replica Analytic Continuation] 
With respect to models where we can assess the FSNJ and FJNS 
	   approaches without approximation, 
for $F(n)=E[Z^n],(n\in{\bf C})$ analyzed by the FSNJ procedure 
and $G(n)=E[Z^n],(n\in{\bf Z})$ analyzed by the FJNS procedure, 
when $n\in{\bf Z}$, 
$F(n)=G(n)$ is always satisfied.
In addition, 
$G(n)$ is regarded as an analytical function of the replica number $n$,
and we assume that $G(n)$ is also described as an analytical function of $n$
when $n\not\in{\bf Z}$, so then if
$F(n)=G(n),(n\not\in{\bf Z})$ holds, the replica analytic continuation 
	   for this model is guaranteed.
\end{description}
Namely, 
given the positive results of our study, it can now be expected that 
any doubts about the accuracy of replica analysis will not be directed at replica 
analytic continuation.
Hence any questions regarding replica analysis reported in previous works are likely to be based on 
more complicated aspects of replica analysis, particularly 
when the thermodynamical limit and the expansion on order parameters are included.
This paper has taken the approach of decomposing replica analysis into its composite components, 
allowing the mathematical properties of each component to be investigated in order to clarify the mathematical structure of replica analysis.
However, it is also important to examine
problems where the components are used
in combination.

\section{Conclusion and Future Works}
In this work, 
to explore the roots of questions on the validity of replica analysis, 
we applied replica analytic continuation to analyze harmonic oscillator models (classical and quantum) and the Ising model, 
which do not need consideration of the thermodynamical limit, and validated the effectiveness of 
replica analytic continuation.
Although the present paper handles only a few examples,  
we discussed whether replica analytic continuation holds 
with respect to toy models for which it is possible to solve 
$F(n)=E[Z^n],(n\in{\bf C})$ and $G(n)=E[Z^n],(n\in{\bf Z})$ using the FSNJ and FJNS approaches without approximations.
Although we cannot provide a universal proof on the validity of replica analysis in this 
paper, {our goal
was} to formulate a conjecture for replica analytic continuation. 
We do not discuss the thermodynamical limit and order parameter 
expansion, though these have already been guaranteed theoretically and practically,
and are not particularly under question.
However, considering more complicated cases will be difficult \cite{Anderson} 
as even if each analytical technique 
can be proved individually, 
they are complicated when combined.

In future work 
we need to discuss whether 
replica analytic continuation holds for 
physical or statistical models of many-body systems using FSNJ 
and FJNS without approximations. 
Although we considered a few positive examples in this paper, 
it is preferable to prove the conjecture of replica analytic continuation in 
a general form. Furthermore, although we discussed 
replica analytic continuation
using the versatile properties of the Gamma function,
other special functions could further support the 
proof of replica analytic continuation. 

\section*{Acknowledgements}
The author appreciates the helpful comments offered by Y. Kabashima,  
 T. Tanaka and K. Kobayashi. 
The work is partly supported by Grant-in-Aid Nos. 24710169 and 15K20999; the
President Project for Young Scientists at Akita Prefectural University; 
grant No. 50 of the National Institute of Informatics, Japan; 
grant No. 5 of the Japan Institute of Life Insurance; 
a grant from the Institute of Economic Research Foundation; Kyoto University; 
grant No. 1414 of the Zengin Foundation for Studies in Economics 
and Finance; 
grant No. 2068 of the Institute of Statistical Mathematics; research project of Mitsubishi UFJ Trust Scholarship Foundation; 
and grant No. 2 of the Kampo Foundation.

\appendix
\section{Calculation of $E[\log Z]$}
Here we discuss the expectation of the logarithm of the partition 
function $E[\log Z]$. It is well-known that the Gamma function of $z$, $\Gamma(z)=\int_0^\infty 
dtt^{z-1}e^{-t}$ is described as
\bea
\Gamma(z)\eq \f{1}{z}e^{-\g z}\prod_{k=1}^\infty
\left(\f{k}{k+z}\right)e^{\f{z}{k}}.
\eea
Using this description, the derivative of the logarithm function of the Gamma 
function with respect to $z$, 
\bea
\psi(z)\eq\pp{}{z}\log\Gamma(z)\nn
\eq-\g-\f{1}{z}+\sum_{k=1}^\infty
\left(\f{1}{k}-\f{1}{k+z}\right),
\eea
is easily calculated where the Euler constant $\g$ is defined as
\bea
\g\eq\lim_{L\to\infty}\left(\sum_{k=1}^L\f{1}{k}-\log L\right)\nn
&\simeq&0.57721.
\eea
We call $\psi(z)$ the digamma function.
Especially for $z\in{\bf Z}$, the digamma function is formulated as follows;
\bea
\psi(z)\eq-\g+\sum_{k=1}^{z-1}\f{1}{k},\\
\psi\left(z+\f{1}{2}\right)\eq-\g-2\log2+\sum_{k=1}^z\f{1}{k-\f{1}{2}}.
\eea
From which we have,
\bea
\psi(1)\eq-\g,\\
\psi\left(\f{1}{2}\right)\eq-\g-2\log2.
\eea
From these settings, the expectation of the logarithm 
of the partition function $E[\log Z]$ in each case is solved as follows:
\subsection{Expectation of $\log Z$ in Section \ref{sec3.1}}
\bea
E[\log Z]
\eq\f{\g}{2}-\f{1}{2}\log 2-\log(\hbar\b),
\eea
\subsection{Expectation of $\log Z$ in Section \ref{sec3.2}}
\bea
E[\log Z]
\eq\f{\g}{2}+\f{1}{2}\log 2-\log(\hbar\b\s),
\eea

\subsection{Expectation of $\log Z$ in Section \ref{sec3.3}}
\bea
E[\log Z]
\eq-\f{1}{2}\psi\left(\f{s}{2}\right)-\f{1}{2}\log 2-\log(\hbar\b),
\eea
\subsection{Expectation of $\log Z$ in Section \ref{sec3.4}}
\bea
E[\log Z]
\eq
\f{1}{2}\psi\left(\f{1}{\hbar\b\mu}\right)+
\f{1}{2}\psi\left(1+\f{1}{\hbar\b\mu}\right)-\psi(1)\nn
\eq-\f{\hbar\b\mu}{2}+\sum_{k=1}^\infty\left(\f{1}{k}-\f{1}{k+\f{1}{\hbar\b\mu}}\right),
\eea
\subsection{Expectation of $\log Z$ in Section \ref{sec3.5}}
\bea
E[\log Z]
\eq
\area Dx\log2\cosh\left(\b\mu+\b\s x\right),\qquad
\eea
\subsection{Expectation of $\log Z$ in Section \ref{sec3.6}}
\bea
E[\log Z]
\eq
-\log (h\b g)-\psi(\tau),
\eea
\subsection{Expectation of $\log Z$ in Section \ref{sec3.7}}
\bea
E[\log Z]
\eq
\log 2+\psi\left(\f{m}{2}\right),
\eea
\subsection{Expectation of $\log Z$ of 1st Example in Section \ref{sec3.8}}
\bea
E[\log Z]
\eq
\f{\s^2}{2},
\eea
\subsection{Expectation of $\log Z$ of 2nd Example in Section \ref{sec3.8}}
\bea
E[\log Z]
\eq
\log\f{t}{s}+\psi\left(\f{s}{2}\right)-\psi\left(\f{t}{2}\right),
\eea

\if 0
\begin{description}
\item[Expectation of $\log Z$ in Section \ref{sec3.1}]
\bea
E[\log Z]
\eq\f{\g}{2}-\f{1}{2}\log 2-\log(\hbar\b),
\eea
\item[Expectation of $\log Z$ in Section \ref{sec3.2}]
\bea
E[\log Z]
\eq\f{\g}{2}+\f{1}{2}\log 2-\log(\hbar\b\s),
\eea

\item[Expectation of $\log Z$ in Section \ref{sec3.3}]
\bea
E[\log Z]
\eq-\f{1}{2}\psi\left(\f{s}{2}\right)-\f{1}{2}\log 2-\log(\hbar\b),
\eea
\item[Expectation of $\log Z$ in Section \ref{sec3.4}]
\bea
E[\log Z]
\eq
\f{1}{2}\psi\left(\f{1}{\hbar\b\mu}\right)+
\f{1}{2}\psi\left(1+\f{1}{\hbar\b\mu}\right)-\psi(1)\nn
\eq-\f{\hbar\b\mu}{2}+\sum_{k=1}^\infty\left(\f{1}{k}-\f{1}{k+\f{1}{\hbar\b\mu}}\right),
\eea
\item[Expectation of $\log Z$ in Section \ref{sec3.5}]
\bea
E[\log Z]
\eq
\area Dx\log2\cosh\left(\b\mu+\b\s x\right),\qquad
\eea

\item[Expectation of $\log Z$ in Section \ref{sec3.6}]
\bea
E[\log Z]
\eq
-\log (h\b g)-\psi(\tau),
\eea
\item[Expectation of $\log Z$ in Section \ref{sec3.7}]
\bea
E[\log Z]
\eq
\log 2+\psi\left(\f{m}{2}\right),
\eea
\item[Expectation of $\log Z$ of 1st Example in Section \ref{sec3.8}]
\bea
E[\log Z]
\eq
\f{\s^2}{2},
\eea
\item[Expectation of $\log Z$ of 2nd Example in Section \ref{sec3.8}]
\bea
E[\log Z]
\eq
\log\f{t}{s}+\psi\left(\f{s}{2}\right)-\psi\left(\f{t}{2}\right),
\eea
\end{description}
\fi


\begin{thebibliography}{99}



\bibitem{Mezard1987}
 \citebook{
 \citeauthorname{M. }{M$\acute{\rm e}$zard},  \citeauthorname{G. }{Parisi}, 
 \citeauthorname{M. A. }{Virasoro}, 
}
{Spin glass theory and beyond%: An Introduction To The Replica Method And Its Applications
}
{World Scientific Publishing}{1987}


\bibitem{Nishimori}
 \citebook{ \citeauthorname{H.}{Nishimori}, }
{Statistical physics of spin glasses and information 
	processing} 
{Oxford University Press}{2001}





\bibitem{Mezard}
 \citebook{
 \citeauthorname{M. }{M$\acute{\rm e}$zard},  \citeauthorname{A. }{Montanari}, }
{Information, Physics, and Computation}
{Oxford University Press}{2009}





\bibitem{EA}
 \citepaper{
 \citeauthorname{S. F. }{Edwards}, 
 \citeauthorname{P. W. }{Anderson}, 
}{Theory of spin glasses}
{%Journal of Physics F: Metal Physics 5(5) 965?974.
J. Phys. F, {\bf 5}, 965}
{1975}

\bibitem{SK}
 \citepaper{
 \citeauthorname{D. }{Sherrington}, 
 \citeauthorname{S. }{Kirkpatrik}, 
}{Solvable model of a spin-glass}
{%Physics Review Letters 35 (26): 1792?1796
Phys. Rev. Let., {\bf 35}, 1792}
{1975}


\bibitem{KS}
 \citepaper{
 \citeauthorname{S. }{Kirkpatrik}, 
 \citeauthorname{D. }{Sherrington}, 
}{Infinite-ranged models of spin-glasses}
{%Physics Review Letters 17(11): 4384-4403
Phys. Rev. B, {\bf 17}, 4384}
{1978}


\bibitem{Amit1}
 \citepaper{
 \citeauthorname{D. J.}{Amit},
 \citeauthorname{H. }{Gutfreund},
 \citeauthorname{H. }{Sompolinsky}, 
}{Spin-glass model of neural networks}
{%Physics Review A 32(2) 1007-1018
Phys. Rev. A, {\bf 32}, 1007}
{1985}

\bibitem{Amit2}
 \citepaper{
 \citeauthorname{D. J.}{Amit},
 \citeauthorname{H. }{Gutfreund},
 \citeauthorname{H. }{Sompolinsky}, 
}{Storing infinite numbers of patterns in a spin-glass model of neural networks}
{%Physics Review Letters 55 (14) 1530-1533 
Phys. Rev. Let., {\bf 55}, 1530}
{1985}

\bibitem{Gardner}
 \citepaper{
 \citeauthorname{E. }{Gardner},
}{The space of interractions in neural network models}
{%Jounal of Physics A 21(1) 257-270
J. Phys. A, {\bf 21}, 257}
{1988}


\if 0
\bibitem{Shinzato2008}
 \citepaper{ \citeauthorname{T.}{Shinzato},
 \citeauthorname{Y.}{Kabashima},
}{Perceptron capacity revisited: 
	classification ability for correlated patterns}
{%The Journal of Physics, A. 41(32),  324013
J. Phys. A. {\bf 41},  324013
}
{2008}
\fi
\bibitem{Kabashima2004}
 \citepaper{ \citeauthorname{Y.}{Kabashima},
 \citeauthorname{D.}{Saad},}
{Statistical mechanics of low-density parity-check codes}
{%The Journal of Physics, A. 37(6), R1-R43.
J. Phys. A. {\bf 37}, R1
}
{2004}

\bibitem{Tanaka2002}
 \citepaper{
 \citeauthorname{T.}{Tanaka},
}{A statistical-mechanics approach to large-system analysis of CDMA multiuser detectors}
{%IEEE Transactions on Information Theory, 48(11), 2888-2910,
IEEE on IT, {\bf 48}, 2888
}
{2002}
\bibitem{Ciliberti}
 \citepaper{
 \citeauthorname{S.}{Ciliberti} 
 \citeauthorname{M.}{M$\acute{\rm e}$zard}, }
{Risk minimization through portfolio replication}
{%The European Physical Journal B, 27(57), 175-180
Euro. Phys. J. B, {\bf 27}, 175}{2007}

\bibitem{Shinzato-SA2015}
 \citepaper{ \citeauthorname{T.}{Shinzato}, }
{Self-averaging property of minimal investment risk of mean-variance 
	model}
{%Public Library of Science One, 10(7), e0133846.
PLoS One, {\bf 10}, e0133846
}{2015}




\bibitem{Kac}
 \citepaper{
 \citeauthorname{M. }{Kac},
}{}
{Unpublished. Trondheim Theoretical Physics Seminar, Nordita Publ., No. 
	286}
{1968}

\bibitem{Lin}
 \citepaper{
 \citeauthorname{T.-F. F. }{Lin}, 
}{Problem of the Disordered Chain}{%Journal of Mathematical Physics, 11(5), 1584-1590
J. Math. Phys. {\bf 11}, 1584
}{1970}

\bibitem{Edwards1970}
 \citepaper{
 \citeauthorname{S. F. }{Edwards},
}{Statistical Mechanics of Polymerized Materials}
{%Proceedings of the 4th International Conference on Amorphous Materials, 
	% 279-300.
Proc. 4th Int. Conf. Amor. Mat., 279, 
{%eds. R. W. Douglas and B. Ellis, 
Wiley, New York}
}
{1970}
\bibitem{Edwards1971}
 \citepaper{
 \citeauthorname{S. F. }{Edwards},
}{Statistical Mechanics of Rubber}
{Polymer Networks: Structural and Mechanical Properties, 
%eds. A. Chompff, S. Newman, 
Plenum Press, New York}
{1971}



\bibitem{Emery}
 \citepaper{
 \citeauthorname{V. J. }{Emery}, 
}{Critical properties of many-component systems}{%Physical Review B, 11(1), 239-247
Phys. Rev. B, {\bf 11}, 239
}{1975}



\bibitem{Jasnow}
 \citepaper{
 \citeauthorname{D. }{Jasnow}, 
 \citeauthorname{M. E. }{Fisher}, 
}{Self-interacting walks, random spin systems, and the zero-component limit}{%Physical Review B, 13(3), 1112-1118
Phys. Rev. B, {\bf 13}, 1112
}{1976}

\bibitem{Grinstein}
 \citepaper{
 \citeauthorname{G. }{Grinstein}, 
 \citeauthorname{A. }{Luther}, 
}{Application of the renormalization group to phase transitions in disordered systems}{%Physical Review B, 13(3), 1329-1343
Phys. Rev. B, {\bf 13}, 1329
}{1976}







\bibitem{Cover}
 \citepaper{
 \citeauthorname{T. M. }{Cover},
}{Geometrical and Statistical Properties of Systems of Linear Inequalities with Applications in Pattern Recognition}
{%IEEE Transactions on Electronic Computers, {\bf 14}(3) 326-334
IEEE Trans. Electron. Comput., {\bf 14} 326
}
{1965}



\bibitem{Venkatesh}
 \citepaper{
 \citeauthorname{S. S. }{Venkatesh},
}{Epsilon capacity of neural networks}
{%American Institute of Physics Conference Proceedings, {\bf 151}(1) 440-445
AIP, Conf. Proc., {\bf 151}, 440}
{1986}




\bibitem{Shinzato-Kabashima}
{\citepaper{
\citeauthorname{T. }{Shinzato}, 
\citeauthorname{Y. }{Kabashima}, 
}
{Perceptron Capacity Revisited: Classification Ability for Correlated Patterns}
{J. Phys.  A, {\bf 41}, 324013}
{2008}
}
\bibitem{AT}
 \citepaper{
 \citeauthorname{J. R. L. }{Almeida},
 \citeauthorname{D. J. }{Thouless}, 
}{Stability of the Sherrington-Kirkpatrick solution of a spin glass model}
{%Journal of Physics A 11(5): 983-990
J. Phys.  A, {\bf 11}, 983}
{1978}



\bibitem{Parisi1}
 \citepaper{
 \citeauthorname{G. }{Parisi},
}{The order parameter for spin glasses: a function on the interval 0-1}
{%Journal of Physics A 13(3): 1101-1112
J. Phys.  A, {\bf 13}, 1101}
{1980}


\bibitem{Parisi2}
 \citepaper{
 \citeauthorname{G. }{Parisi},
}{A sequence of approximated solutions to the S-K model for spin glasses}
{%Journal of Physics A 13(4): L115-L121
J. Phys.  A, {\bf 13}, L115}
{1980}




\bibitem{Ogure1}
 \citepaper{
 \citeauthorname{K. }{Ogure},
 \citeauthorname{Y. }{Kabashima},
}{Exact Analytic Continuation with Respect to the Replica Number in the Discrete Random Energy Model of Finite System Size}
{%Progress of Theoretical Physics Supplement, {\bf 111}(5) 661-688
Prog. Theor. Phys., {\bf 111}, 661}
{2004}

\bibitem{Ogure2}
 \citepaper{
 \citeauthorname{K. }{Ogure},
 \citeauthorname{Y. }{Kabashima},
}{An Exact Analytic Continuation to Complex Replica Number in the Discrete Random Energy Model of Finite System Size}
{%Progress of Theoretical Physics Supplement, {\bf 157} 103-106
Prog. Theor. Phys. Suppl., {\bf 157}, 103}
{2005}


\bibitem{Tanaka2007}
 \citepaper{
 \citeauthorname{T. }{Tanaka},
}{Moment Problem in Replica Method}
{%Interdisciplinary Information Sciences, {\bf 13}(1) 17-23
Interdis. Info. Sci., {\bf 13}, 17}
{2007}

\bibitem{Artin}
 \citebook{ \citeauthorname{E.}{Artin}, }
{The Gamma Function} 
{Dover Publications}{2015}

\bibitem{Andrews}
\citebook{ 
\citeauthorname{G. E. }{Andrews}, 
\citeauthorname{R. }{Askey}, 
\citeauthorname{R. }{Roy},
}
{Special Functions} 
{Cambridge University Press}{2001}

\bibitem{Levedev}
 \citebook{ \citeauthorname{N. N. }{Levedev}, }
{Special Functions \& Their Applications}
{Dover Publications}{1972}

\bibitem{Anderson}
\citepaper{
\citeauthorname{P. W. }{Anderson},
}{More is different}
{%Science, 177, 393-396, 1972
Science, {\bf 177}, 393}{1972}


\end{thebibliography}
\end{document}